\documentclass[aps, prl, twocolumn, showpacs, superscriptaddress]{revtex4}
\usepackage{graphicx}

\begin{document}

\title{Ultrasonic triggering of giant magnetocaloric effect in MnAs thin films.}

  \author{J.-Y. Duquesne}
  \affiliation{Institut des
  NanoSciences de Paris, UPMC-CNRS UMR 7588, 4 place Jussieu, 75252
  Paris Cedex 5, France.}
 \author{J.-Y. Prieur}
 \affiliation{Institut
  des NanoSciences de Paris, UPMC-CNRS UMR 7588, 4 place Jussieu,
  75252 Paris Cedex 5, France.} 
 \author{J. Agudo Canalejo}
 \affiliation{Institut des NanoSciences de Paris, UPMC-CNRS UMR 7588, 4
  place Jussieu, 75252 Paris Cedex 5, France.}
 \author{V.H. Etgens}
 \affiliation{Institut des NanoSciences de Paris,
  UPMC-CNRS UMR 7588, 4 place Jussieu, 75252 Paris Cedex 5,
  France.}
 \affiliation{F\'{e}d\'{e}ration Lavoisier Franklin, UVSQ, 45
  avenue des Etats Unis - 78035 Versailles cedex, France.}
 \author{M. Eddrief}
 \affiliation{Institut des NanoSciences de Paris,
  UPMC-CNRS UMR 7588, 4 place Jussieu, 75252 Paris Cedex 5, France.}
 \author{A.L. Ferreira}
 \affiliation{Institut des NanoSciences de Paris,
  UPMC-CNRS UMR 7588, 4 place Jussieu, 75252 Paris Cedex 5,
  France.}
 \affiliation{Departamento de F\'isica, UFPR, Centro
  Polit\'ecnico, Caixa Postal 19091, 81531-990, Curitiba PR, Brazil.}
 \author{M. Marangolo}
 \affiliation{Institut des NanoSciences de Paris,
  UPMC-CNRS UMR 7588, 4 place Jussieu, 75252 Paris Cedex 5, France.}

\date{\today}

\begin{abstract}
Mechanical control of magnetic properties in magnetostrictive thin
films offers the unexplored opportunity to employ surface wave
acoustics in such a way that acoustic triggers dynamic
magnetic effects. The strain-induced modulation of the magnetic
anisotropy can play the role of a high frequency varying effective
magnetic field leading to ultrasonic tuning of electronic and magnetic
properties of nanostructured materials, eventually integrated in
semiconductor technology. Here, we report about the opportunity to
employ surface acoustic waves to trigger magnetocaloric effect in
MnAs(100nm)/GaAs(001) thin films. During the MnAs magnetostructural
phase transition, in an interval range around room temperature
(0$^{\circ}$C~-~60$^{\circ}$C), ultrasonic waves (170 MHz) are
strongly attenuated by the phase coexistence (up to 150~dB/cm). We
show that the giant magnetocaloric effect of MnAs is responsible of
the observed phenomenon. By a simple anelastic model we describe the
temperature and the external magnetic field dependence of such a huge
ultrasound attenuation. Strain-manipulation of the magnetocaloric
effect could be a further interesting route for dynamic and static
caloritronics and spintronics applications in semiconductor
technology.
\end{abstract}


\maketitle

%%%%%%%%%%%%%%%%%%%%%
\section{Introduction}
In the last few years, magnetism research in bulk materials and in
nanostructures cross-coupled magnetization with either local and
non-inductive fields or with thermally driven effects. Former
experiments concern electric field to control local magnetization in
multiferroic materials \cite{ref:Chu}, spin polarized currents to
generate RF coherent emission in nanopillars \cite{ref:Berger},
ultrafast pulsed lasers to create magnetic domains \cite{deJong}, self
organized templates to switch magnetization \cite{ref:Sacchi}. Latter
experiments deal with a combination of electron spin and heat, such as
spin-dependent Peltier \cite{ref:Gravier} and Seebeck effects
\cite{ref:Uchida}. All these effects would permit to obtain new means
to control local magnetic properties in spintronics devices avoiding
cumbersome inductive means.
Here, we report on a thermally driven effect induced by non-inductive
means in a magnetic thin film.  We show that the well-known
interaction of surface acoustic waves (SAWs) with magnetic excitations
\cite{ref:Davis, ref:Weiler, ref:Huber} is able to trigger MCE in
MnAs. We argue that strain induces strong modifications of the inner
magnetic field in this magnetoelastic material inducing consequent MCE
triggering.

\section{Magnetocaloric properties of M\lowercase{n}A\lowercase{s}}
The magnetocaloric effect (MCE) is either an isothermal
magnetic-entropy change or an adiabatic temperature change, obtained
by applying an external magnetic field. MCE enables an efficient
refrigeration process and would permit a low cost and environment
sustainable alternative to gas compression techniques
\cite{ref:Gschneidner}. Recently, our group has shown that MCE can be
obtained in MnAs thin films and tailored by epitaxial strain
engineering \cite{ref:Mosca2008}. MnAs presents also a strong
magnetoelastic coefficient that induces a magnetostructural
transition around 40$^\circ$C in bulk MnAs, where the low temperature
hexagonal NiAs structure ($\alpha$-MnAs) transforms into
non-ferromagnetic orthorhombic $\beta$-MnAs by a first order
transition \cite{ref:Bean}. This magnetostructural phase transition is
accompanied by one of the highest magnetocaloric effect, in the
neighboring of room temperature \cite{ref:vonRanke}:
120~J~Kg$^{-1}$K$^{-1}$ for a magnetic field variation of 5 T. In
MnAs/GaAs(001) thin films, magnetoelastic coupling is so intense that,
over the temperature range 13-40 $^\circ$C, MnAs/GaAs(001) displays an
$\alpha / \beta$-MnAs phase coexistence \cite{ref:Kaganer,
  ref:Breitwieser2009}. This phenomemon is due to epitaxial conditions
\cite{ref:Daeweritz}. The $\alpha / \beta$ phase coexistence induces a
spreading of the magnetocaloric effect in a temperature range,
centered around 35$^\circ$C \cite{ref:Mosca2008} . The magnetic
entropy change depends quite linearly on the applied magnetic field
(see figure 3 in ref.\cite{ref:Mosca2008}).

%%%%%%%%%%%%%%%%%%%%%%%%%%%%%%%%%%%%%%%%%%%%%%%%%%%%%%%%
%%  Manips
%%%%%%%%%%%%%%%%%%%%%%%%%%%%%%%%%%%%%%%%%%%%%%%%%%%%%%%%
\section{Experimental methods}
\begin{figure}
  \includegraphics[width=0.95\columnwidth]{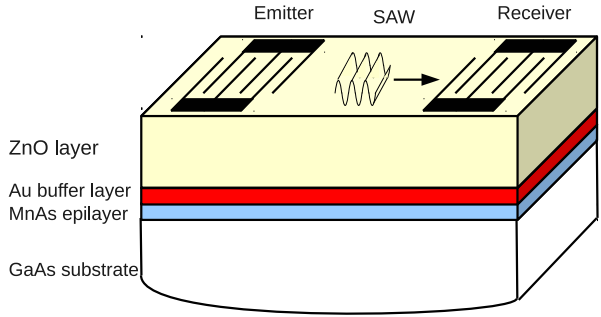}
  \caption{(color online) Structure of the sample (not to scale). MnAs
    epilayer: 100 nm. Gold buffer: $\simeq 300$ nm. ZnO piezoelectric
    layer $\simeq 1 \mu$m. Excited wavelength: $\lambda = 15
    \mu$m. The acoustic wave vector is parallel to MnAs easy magnetic
    axis.}
    \label{fig:SAW}
\end{figure}

Figure \ref{fig:SAW} displays the structure of the sample under study.
MnAs epilayers were grown by MBE on GaAs(001) substrates.  Epiready
GaAs substrates were first deoxidized under As overpressure followed
by a GaAs buffer-layer growth in standard growth conditions. At the
end, the surface was long annealed at 600 $^\circ$C under As to
optimize its quality, confirmed by the presence of a clear
$(2 \times 4)\beta$ diagram as checked by reflection high-energy electron
diffraction (RHEED). Next, we have cooled down the sample and followed
the procedure of Arai {\it et al} \cite{ref:Arai2007} to obtain a stable and
high-quality As-terminated c~(4x4) surface. The MnAs growth was
performed at 260$^\circ$C under As-rich conditions and a growth rate
of about 3 nm/min. The epitaxial relationship was first verified in
situ by RHEED and crosschecked ex situ by x-ray diffraction. MnAs
displayed a single domain epitaxy from the beginning of the growth
with [0001]MnAs//[1-10]GaAs. Finally samples were protected by few
nanometers Au capping layer. The sample is then introduced in a
sputtering chamber where a gold layer ($\sim$200~nm) and a ZnO layer
($\sim$2~$\mu$m) are deposited. The substrate temperature is around
200$^{\circ}$C. The gold layer is obtained by thermal evaporation. The
piezoelectric ZnO layer is obtained by RF sputtering from a Zn target
in an Ar+O$_{2}$ plasma. The aim of the gold layer is to favor ZnO
growth and to relax the constraints arising from the different
dilatations of MnAs/GaAs and ZnO. The aim of the piezoelectric ZnO layer
is to excite and detect surface acoustic waves, thanks to interdigital
transducers (IDTs).  IDTs are oriented in such a way the acoustic
wavevector is parallel to the easy magnetic axis of MnAs
$\alpha$-phase. ITDs are made by lift-off photolithography and thermal
evaporation of gold (200 nm) on a thin adhesion Cr layer. The period
of the IDTs is 15$\mu$m. The width of one IDTs tooth is
3.75$\mu$m. The excited wavelength is $15~\mu$m and the corresponding
resonant acoustic frequency is 170 MHz. The distance between
transducers is 2~mm. The aperture of the transducers is 2~mm. The
emitter is excited at its resonant frequency (170 MHz) with a 500 ns
RF burst. After propagation in the sample, the acoustic burst is
detected by the receiver and the signal is processed using a phase
detection scheme. Because of the velocity changes versus temperature
or magnetic field, the resonant frequency of the transducers slightly
shifts with temperature or field. However, we checked that the shift
is small and has no affect on the measured attenuation variation, at
fixed frequency. Typical cooling and warming rate is 1$^{\circ}$C per
minute.

\section{Experimental resultats}
\begin{figure} 
  \includegraphics[angle=0, width=0.95\columnwidth]{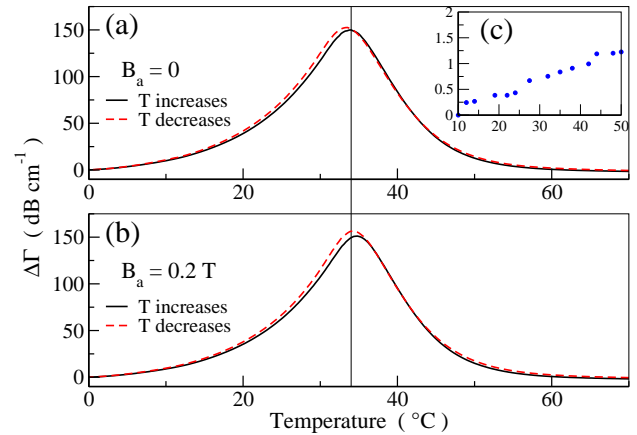}
  \caption{(color online) Variations of acoustic attenuation $\Delta
    \Gamma = \Gamma(T) - \Gamma(0^\circ$C) versus temperature $T$, at
    170 MHz, for two different applied magnetic fields. Inset:
    Attenuation changes $\Delta \Gamma = \Gamma(T) -
    \Gamma(10^\circ$C) in the MnAs free sample, at 120 MHz.}
  \label{fig:att_T}
\end{figure}
In a first set of experiments, the sample is demagnetized at
$86^\circ$C and zero-field-cooled down to $0^\circ$C. The attenuation
changes are then measured as a function of temperature, from
$0^\circ$C to $86^\circ$C, and back to $0^\circ$C. Thermal cycles are
performed either in zero or non-zero applied field (0.2 T). Figure
\ref{fig:att_T} displays the measured difference of ultrasound
attenuation, $\Delta \Gamma$, with respect to a reference value at
$T=0^\circ$C.  The most important experimental result of this paper
is the huge attenuation peak (150 dB~cm$^{-1}$) observed around
$34^\circ$C.  The magnitude of this variation is surprising if one
considers that the MnAs layer (100 nm) is roughly two orders of
magnitude smaller that the penetration depth of the surface acoustic
wave ($\sim \lambda = 15~\mu$m), so that most of the acoustic energy is
located in the substrate, not in the MnAs layer.  The precise location
of the peak slightly depends on whether the sample is warmed or
cooled. Moreover the attenuation curve is rigidly shifted to higher
temperatures when an external field is applied:
\begin{equation}
\label{eq:gammaA}
  \Delta T = \gamma_{A} B_{a}
\end{equation}
where $\gamma_{A}\simeq$4.6$^{\circ}$CT$^{-1}$. For comparison, we
measured the attenuation changes versus temperature in similar hybrid
structure, but without MnAs. Figure \ref{fig:att_T} inset displays the
result. No attenuation peak is observed and the attenuation variation
is only around 1 dB~cm$^{-1}$ between 10 and 50$^\circ$C. 

\begin{figure}
  \includegraphics[angle=0,width=0.95\columnwidth]{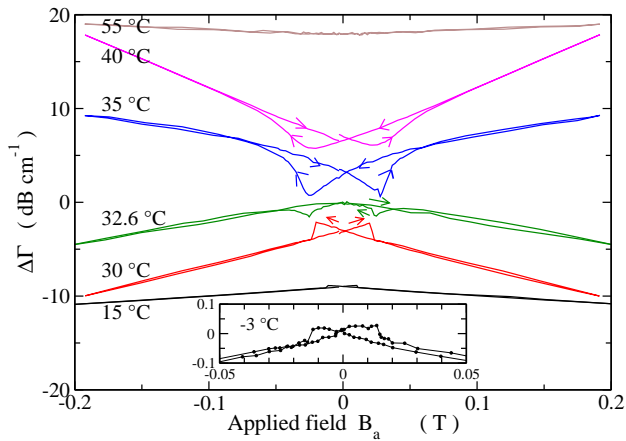}
  \caption{(color online) Isothermal attenuation variations $\Delta
    \Gamma = \Gamma(B_{a}) - \Gamma(B_{a}=0)$, at 170 MHz, versus
    applied field. For clarity, curves are shifted along the
    attenuation axis. The arrows show the direction of the magnetic
    cycle. Inset: $\Delta \Gamma$ at T=$-3^\circ$C.}
  \label{fig:att_H}
\end{figure}
\begin{figure}
  \includegraphics[angle=0,width=0.95\columnwidth]{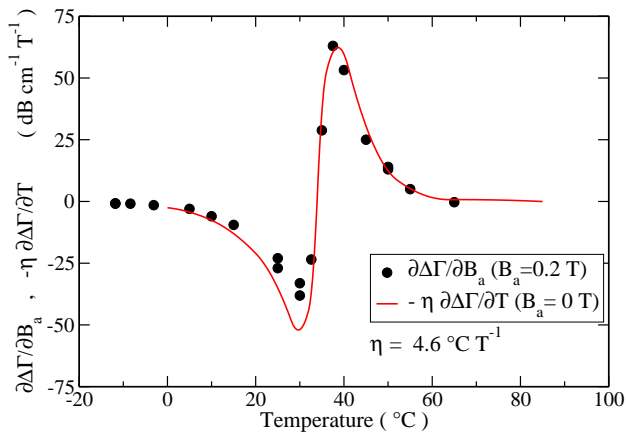}
  \caption{(color online) Temperature  and field derivatives
    of the experimental acoustic attenuation.}
 \label{fig:d_att}
\end{figure}
In a second set of experiments, we measured the isothermal attenuation
versus applied magnetic field. In that case, at every temperature, the
sample is first magnetically cycled before the acoustic measurements
are performed (cycle: 0 to 0.2~T to -0.2~T to
0~T). Figure~\ref{fig:att_H} displays typical results. Hysteresis is
observed at low fields but the overall behavior is a linear variation
of attenuation versus applied field modulus.  It is worthwhile
noticing the striking correlation between the behaviors of attenuation
$\Delta \Gamma$, versus either temperature $T$ or applied field
$B_{a}$. Below $33^\circ$C, $\Delta \Gamma$ increases either if $T$
increases or $B_{a}$ decreases. Above $33^\circ$C, the reverse
behavior is observed. The temperature and field derivatives are
roughly proportional, as shown in figure \ref{fig:d_att}:
\begin{equation}
  \frac{\partial \Delta\Gamma}{\partial B_{a}} \simeq - \eta
  ~\frac{\partial \Delta \Gamma}{\partial T}
  \label{eq:eta}
\end{equation}
We find $\eta = 4.6^{\circ}$CT$^{-1} = \gamma_{A}$. It can be
shown that this a direct consequence of eq.({\ref{eq:gammaA}).

%%%%%%%%%%%%%%%%%%%%%%%%%%%%%%%%%%%%%%%%%%%%%%%%%%%%%%%%%%%%%%%%%%%%%%%
\section{Discussion}

We argue here that the behavior of the acoustic wave is due to a
thermoelastic effect, enhanced by the magnetoelastic and
magnetocaloric properties of MnAs. These two points will now be
successively discussed.

%%%%%%%%%%%%%%%%%%%%%%%%%%%%%%%%%%%%%%%%%%%%%%%%%%%%%%%%%%%%%%%%%%%%%%%
% \section {Att\'enuation du son par l'effet thermo\'elastique.}
Let us first consider the attenuation of sound due to a relaxation
process.  Quite generally, it can be inferred from a quasi-static
stress-relaxation experiment where a strain $\epsilon$ is suddently
applied and held constant \cite{ref:Nowick}. The stress $\sigma$
instantaneously changes from 0 to $C_{U}\epsilon$ and then relaxes to
$C_{R}\epsilon$ with a characteristic time $\tau$. $C_{U}$ and $C_{R}$
are the instantaneous and relaxed elastic modulus, respectively, and
it is useful to define $\Delta C = C_{U}-C_{R}$. The sound amplitude
decay $\alpha$, defined by the displacement change $u(x)=u(0)\exp
(-\alpha x)$ along the $x$ spatial coordinate, can be inferred
\cite{ref:Nowick}:
\begin{equation}
  \label{eq:att}
      \alpha = \frac{\Delta C}{2 \rho v^{3}}
               \frac{\omega^{2} \tau}{1 + \omega^{2} \tau^{2}}    
\end{equation}
$\omega/2\pi$, $v$ and $\rho$ are the acoustic frequency, the wave
velocity and mass density, respectively. The acoustic attenuation
expressed in dB per unit length is $\Gamma = (20\alpha/\ln 10) \simeq
8.7\alpha$. The expression of $\Delta C$ depends on the relaxation
mechanism. Thermoelastic relaxation is a well known process for sound
absorption due to heat transfer between regions exhibiting different
strains. In that case, the entropy density is the
internal variable which relax upon application of
strain \cite{ref:Nowick}. Its equilibrium value $s$ depends on the
state of strain and the equilibrium time is $\tau$. It can be shown
that
\begin{equation}
   \Delta C = \frac{-1}{\alpha_{s}}
   \left( \frac{\partial s}{\partial \epsilon} \right)_{T}
   \label{eq:DM}
\end{equation}
where $\alpha_{s}$ is the linear thermal expansion coefficient, at
constant entropy. The thermal expansion factor $\alpha_{s}$ is weakly
temperature dependent because MnAs average lattice parameter follows
the thermal lattice expansion of the GaAs substrate
\cite{ref:GarciaPRL}. We assume that $\tau$ arises from heat
transfer from MnAs to the GaAs substrate and Au/ZnO layers, and to the
heat diffusion inside those media. It is estimated to be around
10$^{-11}$s and is also weakly temperature
dependent \cite{ref:consideration}. Consequently, we get $\omega \tau
<< 1$ and
\begin{equation}
  \alpha =  \frac{-1}{2 \rho v^{3} \alpha_{s}} \left(\frac{\partial s}{\partial \epsilon} \right)_{T} \omega^{2} \tau
  \label{eq:alpha}
\end{equation}
Therefore, $(\partial s/\partial \epsilon )_{T}$ is the leading
parameter which governs the thermoelastic attenuation versus
temperature or applied fied. An expression which will be useful later
comes from the Maxwell thermodynamical relation:
\begin{equation}
  \left( 
  \frac{\partial s}{\partial \epsilon}
  \right)_{T}
  =
  -\left( 
  \frac{\partial \sigma}{\partial T}
  \right)_{\epsilon}  
\label{eq:Maxwell}
\end{equation}

Let us now consider the magnetic properties.  In a magnetostrictive
material, acoustic strain field induces a
modification of the internal magnetic field. Here, we propose
that this field modulation triggers a dynamic magnetocaloric effect in
MnAs, responsible for large entropy and heat production. This enhances
thermoelastic attenuation.  The following
experimental and theoretical observations suggest a magnetic excitations {\it scenario}:

(i) The open hysteresis of figure \ref{fig:att_H} attests that we deal
with a magnetic phenomenon determined by the MnAs thin film. It is
worthwhile emphasizing the measured coercitive fields are in good
agreement with the values derived from independant Kerr measurements
(not shown here). Moreover, no large attenuation changes are observed
in a similar sample, without MnAs layer (see fig. \ref{fig:att_T} inset).

(ii) We underline the strong analogies between the attenuation and the
field induced magnetic entropy changes $\Delta s_{m} = s_m(B, T) -
s_m(0, T)$ which has been measured in MnAs thin films by Mosca {\it et
  al} \cite{ref:Mosca2008}. Both exhibit an extrema value and are
symmetric with respect to $\simeq 33^{\circ}$C. Both exhibit high
values at high temperatures ($\simeq 45^\circ$C) and are magnetic
field dependent (figure \ref{fig:att_H}), despite the low
$\alpha$-MnAs fraction value. No saturation trend is observed at
higher fields.

(iii) It has been shown that $\Delta s_{m}$ is approximately equal to the
total entropy change at the MnAs phase transition,
indicating the important role of the magnetocaloric effect \cite{Zou}.

Now we show, qualitatively and quantitatively, that the magnetoelastic and
magnetocaloric properties of MnAs thin films can explain the behavior
of surface acoustic waves that we have observed.

%%%%%%%%%%%%%%%%%%%%%%%%%%%%%%%%%%%%%%%%%%%%%%%%%%%%%%%%%%%%%%%%%
% \section {Das et al. / Maxwell}
\begin{figure}
  \includegraphics[angle=0,width=0.95\columnwidth]{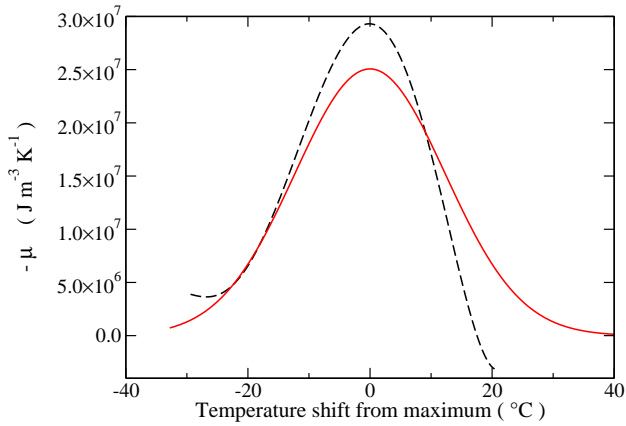}
  \caption
      {(color online) \underline{Dashed line:} $ \mu = \left( \partial
        s / \partial \epsilon \right)_{\epsilon = 0}$ derived from Das
        {\it et al.} \cite{ref:Das2003}.  \underline{Continuous line:}
        $\left( \mu = \partial \Delta s_{m} / \partial \epsilon
        \right)_{\epsilon = 0}$ derived from Mosca {\it et al.}
        \cite{ref:Mosca2008}.  Temperatures of the maxima differ in
        3$^{\circ}$C, due to different samples or experimental
        methods.}
  \label{fig:mu}
\end{figure}
Obviously, in eq.(\ref{eq:DM}, \ref{eq:alpha}, \ref{eq:Maxwell}), $s$
is the total entropy density. It includes the structural entropy of
the substrate as well as the averaged value of the structural and
magnetic entropy $s_{\alpha}$ and $s_{\beta}$ of the pure $\alpha$ and
$\beta$ phases. To get $\Delta C$ from the strain dependence of each
term would be a heroic task. We choose to adopt a pragmatic procedure
by applying the Maxwell relation (\ref{eq:Maxwell}) to the temperature
dependent stress measurements published by Das {\it et al}
\cite{ref:Das2003}. These authors measured the temperature dependence
of the total stress along the a-axis of MnAs thin films. Dashed line
in figure \ref{fig:mu} reports the temperature derivative curve that
we extracted from their published data, i.e. $\left( \partial s /
\partial \epsilon \right)_{T}$. The reader has to take into account
that these measurements were done on MnAs thin films presenting a
slighty lower transition temperature range as compared to our sample.

% \section {Mosca et al. / Magnetic entropy}
To get a more direct insight into the variations of entropy induced by
strain, we consider magnetocaloric measurements. In
\cite{ref:Mosca2008}, the magnetic entropy change as a function of
temperature and magnetic field was extracted from magnetization
versus temperature measurements. The magnetic entropy change follows
the empirical expression between 15 and 55$^{\circ}$C \cite{ref:empiric}:
\begin{equation}
   \Delta s_{m}(B_{a}, T) = - c B_{a}
        ~ \exp
             - \left(
                    \frac{\Delta T- \gamma B_{a}}{d}
             \right)^{2}     
\label{eq:S}
\end{equation}
$B_{a}$ is the applied magnetic field (magnetic flux density),
parallel to the easy magnetic axis of MnAs.  $T$ is the
temperature. $\Delta T = ( T - T_{0} )$ where $T_{0}$ is a reference
temperature. The fitting parameters are $c\simeq2.82 \times10^{4}$
J~m$^{-3}$K$^{-1}$T$^{-1}$, $T_{0}\simeq33^{\circ}$C,
$d\simeq17.4^{\circ}$C, $\gamma\simeq1.8^{\circ}$CT$^{-1}$.  (We
converted the entropy data units from J~kg$^{-1}$K$^{-1}$ to
J~m$^{-3}$K$^{-1}$, using MnAs mass density 6300 kg~m$^{-3}$).  We
notice the entropy curve shifts with $B_{a}$.  Below in the text, we
will show that this $\Delta s_{m}$-shift leads to the attenuation peak
shift that we have observed (4.6$^{\circ}$CT$^{-1}$).  It also recalls
the $\alpha$-fraction shift versus magnetic field observed by Iikawa
{\it et al.} \cite{ref:Iikawa2005a}, with the same order of magnitude
(5$^{\circ}$CT$^{-1}$). The same group reported a similar shift equal
to $\delta \epsilon$ where $\epsilon$ is the strain applied along the
MnAs easy axis, and $\delta$ is temperature independant between 30 and
40$^{\circ}$C (see fig.3 in Ref.\cite{ref:Iikawa2005b}) with $\delta
\simeq 1600^{\circ}$C. Thus, we postulate that an applied strain
$\epsilon$ is equivalent to an effective field $B_{\epsilon} =
(\delta/\gamma)\epsilon$ where $\delta/\gamma \simeq 900$~T.
Consequently, the strain dependance of $\Delta s_{m}$ is merely
obtained by replacing $B_{a}$ with $B_{a}+(\delta/\gamma)\epsilon$ in
eq.(\ref{eq:S}):
\begin{equation}
  \Delta s_{m}(B_{a}, T, \epsilon) = 
    - c (B_{a}+\frac{\delta}{\gamma}\epsilon) 
     ~ \exp
             - \left(
                    \frac{\Delta T- \gamma B_{a} - \delta \epsilon}{d}
             \right)^{2}            
\end{equation}

Using this relation, $\left( \partial
\Delta s_{m} / \partial \epsilon \right)_{T, \epsilon = 0 }$ can be
readily derived:
\begin{eqnarray}
\label{eq:mu2}
  \left( \frac{\partial \Delta s_{m}}{\partial \epsilon} \right)_{T, \epsilon = 0 } =
  \\*
  \nonumber
  -\frac{\delta}{\gamma} c
  \left[
     1 + \frac{2 \gamma }{d^{2}} B_{a} \left( \Delta T -\gamma B_{a}  \right)
  \right]
             \exp
             - \left(
                    \frac{\Delta T- \gamma B_{a}}{d}
             \right)^{2}    
\end{eqnarray}

Figure \ref{fig:mu} displays $\left( \partial \Delta s_{m}/\partial
\epsilon \right)_{T, \epsilon = 0 }$ derived from (\ref{eq:mu2}), at
zero applied field ($B_{a}=0$). Direct comparison can be done with
$\left( \partial s/\partial \epsilon \right)_{T, \epsilon = 0 }$
derived previously from Das {\it et al.} results \cite{ref:Das2003}. We
notice that the magnitude and width of both curves are very close,
despite the different samples and experimental apparatus. {\it This
demonstrates that the main contribution to strain-induced entropy
change comes from the magnetic entropy term.} Consequently, ultrasonic
attenuation $\alpha$ (eq.({\ref{eq:alpha})) is such that:
\begin{equation}
 \alpha
 \propto
 \left( \frac{\partial s}{\partial \epsilon} \right)_{T}
 \simeq
 \left( \frac{\partial \Delta s_{m}}{\partial \epsilon} \right)_{T}
 \label{eq:Dsmag}
\end{equation}
This result is consistent with the aforementioned theoretical
calculations by Zou {\it et al} \cite{Zou}. Strain modifies the
equilibrium state of MnAs inducing MCE, heat flow and anelastic
ultrasonic attenuation which goes with.

 A number of features can be theoretically derived from
 eq.(\ref{eq:alpha}), (\ref{eq:mu2}) and (\ref{eq:Dsmag}) which
 corroborates the {\it scenario} we propose:

{\bf(i) Ultrasonic attenuation order of magnitude:} 

By using eq.(\ref{eq:alpha}), we can roughly estimate the magnitude of
the attenuation peak. Indeed, this equation is valid for plane
waves. In our case, the magnetocaloric energy transfered through the
MnAs/GaAs or MnAs/Au/ZnO interfaces is diluted in the penetration
depth. Thus, we have to normalize eq.(\ref{eq:alpha}) by a geometrical
factor $e/\lambda=0.1 \mu m / 15 \mu m$ where $e$ is the MnAs film
thickness. By using GaAs parameters \cite{ref:GaAs}, we estimate
$\Delta\Gamma = 8.7 \alpha \simeq 150$~dB cm$^{-1}$ at the maximum, in
good agreement with the experimental result. This strongly supports
that MCE in such a thin MnAs film can induce a huge effect on SAWs,
even if the perfect agreement has to be considered as fortuitous
because of the roughness of our approach.

{\bf(ii) Ultrasonic attenuation field dependance:}

Using eq.(\ref{eq:mu2}) and keeping the leading terms, we derive:
\begin{eqnarray}
\label{eq:dAdB}
  \frac {\partial \alpha}{\partial B_{a}}
  \propto
  \frac{\partial}{\partial B_{a}} \left( \frac{\partial \Delta s_{m}}{\partial \epsilon}\right)_{\epsilon = 0}
  = \\*
  \nonumber
  -\frac{4 c \delta}{d^{2}}
  \left(
  \Delta T - \frac{3}{2}\gamma B_{a} 
  + g(\Delta T, B_{a})
  \right)
  \exp
     - \left(
       \frac{\Delta T- \gamma B_{a}}{d}
     \right)^{2}    
\end{eqnarray}
where $g(\Delta T, B_{a}) = \gamma B_{a} (\Delta T - \gamma B_{a})^{2}/d^{2}$.
Eq.(\ref{eq:dAdB}) shows that, in our experimental field range ($B_{a}
< 0.2$ T) and for $\Delta T$ larger than $\gamma B_{a} = 0.36^{\circ}$C,
$\partial \alpha / \partial B_{a}$ is independant of
the field: acoustic attenuation $\alpha$ is a linear function of the
applied field, in good agreement with experiments (figure
\ref{fig:att_H}). 
From eqs.(\ref{eq:mu2}) and (\ref{eq:dAdB}), we can safely
do the following approximation: $(\partial \alpha / \partial
B_{a})/\alpha \simeq (4\gamma/d^{2})\Delta T$. Theoretical and experimental
values are in the same order of magnitude, which is quite
satisfactory, owing to the roughness of our model: 0.2 and 0.6
dB~cm$^{-1}$T$^{-1}$, respectively.

{\bf(iii) Peak temperature dependance:}

Using eq.(\ref{eq:mu2}) and keeping the leading terms, we derive:
\begin{eqnarray}
\label{eq:dAdT}
  \frac {\partial \alpha}{\partial T}
  \propto
  \frac{\partial}{\partial T} \left( \frac{\partial \Delta s_{m}}{\partial \epsilon}\right)_{\epsilon = 0}
  = \\*
  \nonumber
  \frac{2 c \delta}{\gamma d^{2}} \left( \Delta T - 2\gamma B_{a}  + 2 g(\Delta T, B_{a}) \right)
  \exp
     - \left(
       \frac{\Delta T- \gamma B_{a}}{d}       
     \right)^{2}  
\end{eqnarray}

Using eq.(\ref{eq:dAdT}), it is straightforward to show that a good
approximation for the maximum location is given by : $\left( \Delta T
- 2\gamma B_{a} \right)=0$.  Then, our calculation predicts a $2\gamma
= 3.6$~K$^{-1}$T$^{-1}$ attenuation shift with applied field. The
experimental shift is $\gamma_{A}= 4.6$~K$^{-1}$T$^{-1}$. Again, owing
to the roughness of our model, the agreement is quite good.

Finally, it is worthwhile reporting that our model is able to predict
eq.(\ref{eq:eta}), and therefore to establish a connection between
the field induced temperature shifts of both ultrasonic attenuation
and entropy changes.  Indeed, using eq.(\ref{eq:alpha}) and
eqs.(\ref{eq:dAdB}-\ref{eq:dAdT}), we get (in our experimental field
range and for $\Delta T$ larger than a few degrees):
\begin{equation}
  \frac{\partial \alpha}{\partial B_{a}}
  \simeq
  -2\gamma\frac{\partial \alpha}{\partial T}
\end{equation}
A comparison with eq.(\ref{eq:eta}) gives $\eta = \gamma_{A} = 2
\gamma$.

%%%%%%%%%%%%%%%%%%%%%%%%%%%%%%%%%%%%%%%%%%%%%%%%%%%%%%%%%%%%%%%%%%%%%%%%%%
\section{Conclusion}
In conclusion, ultrasound surface waves are a hitherto unexplored mean
to modulate thermal properties in magnetocaloric thin
films, at high frequency, without using cumbersome and
energy-consuming inductive means. We have shown that the huge
ultrasound attenuation observed on thin MnAs films epitaxied on
GaAs(001) can be ascribed to the strain induced triggering of the
giant magnetocaloric effect of MnAs.  In short, we have shown that
acoustic stresses induce local thermodynamical changes in the MnAs and
GaAs media. This changes have a magnetocaloric origin in MnAs despite
the very low quantity of magnetic material in the heterostructure. To
restore equilibrium, a heat flow sets up. Heat transfert is very
rapid.  We anticipate that piezoelectric technology opens up a new way
to control magnetocaloric and spin caloric phenomena, even at high
frequency, in ferromagnet/piezoelectric/semiconductor hybrid
systems. Applications to non-volatile magnetic storage functionality,
fast signal processing, thermal sensors, magnetic sensors and
microwave filtering can be envisaged.

\section{Acknowlegments}
The authors thank M. Sacchi, F. Vidal, D.H. Mosca for fruitful
discussions. A.L. Ferreira acknowledges support from
CAPES-COFECUB. We warmly thank R.Gohier for technical assistance,
M. Brossard and S. Vercauteren for their involvements during their
master internships.

%################################################################
%                       REFERENCES
%################################################################
%% \bibliographystyle{apsrev4-1}
%% \bibliography{MnAs_2.bib}

\begin{thebibliography}{10}
\bibitem{ref:consideration}%
  \BibitemOpen
  \bibinfo {note} {A given amount of thermal energy is driven in the thin film.
  We solve then the heat diffusion equation to monitor the heat transfert rate
  $\tau$ in the substrate. It turns out that $\tau =e^{2}\rho _{f}^{2}C_{f}^{2}
  D_{s} / K_{s}^{2} \simeq 10^{-11}$ s, where $e$, $\rho _{f}$, $C_{f}$ are the
  thickness, mass density and specific heat of MnAs, $D_{s}$, $K_{s}$ are the
  thermal diffusivity and conductivity of GaAs.}%
  \bibAnnoteFile{Stop}{ref:consideration}%
\bibitem{ref:empiric}%
  \BibitemOpen
  \bibinfo {note} {The gaussian fit is a rough approximation of $\Delta s_{m}$
  for $B_{a}<2T$, but it catches the salient physics: maxima vs temperature,
  width of the curves, maximum amplitude vs field and maximum temperature shift
  versus field.}%
  \bibAnnoteFile{Stop}{ref:empiric}%
\end{thebibliography}
%Merlin.mbs v4.21 2009-07-09.
%

%%%%%%%%%%%%%%%%%%%%%%%%%%%%%%%%%%%%%%%%%%%%%%%%%%%%%%%%%%%%%%%%%%%%%%%%%%%%%%
%%%%%%%%%%%%%%%%%%%%%%%%%%%%%%%%%%%%%%%%%%%%%%%%%%%%%%%%%%%%%%%%%%%%%%%%%%%%%%
\end{document}